\documentstyle[prl,aps,multicol]{revtex}

\title{Novel Bell's inequalities for entangled $K^0 \bar{K^0}$ pairs}

\author{A. Bramon$^{1}$ and G. Garbarino$^{2}$}

\address{$^1$Grup de F{\'\i}sica Te\`orica,
Universitat Aut\`onoma de Barcelona, E--08193 Bellaterra, Spain}
\address{$^2$Departament d'Estructura i Constituents de la Mat\`{e}ria,
Universitat de Barcelona, E--08028 Barcelona, Spain} \date{\today}

\begin{document}
\draft
\maketitle

\begin{abstract}
We derive new Bell's inequalities for entangled $K^0\bar{K^0}$ pairs.
This requires 1) mutually exclusive setups allowing either 
$K^0$ {\it vs} $\bar{K^0}$ or $K_S$ {\it vs} $K_L$ detection and 
2) the use of kaon regenerators. The inequalities turn out to
be significantly violated by Quantum Mechanics, resulting in
interesting tests of Local Realism at $\phi$--factories and $p\bar{p}$ machines.
\end{abstract}


\begin{multicols}{2}
 
The correlations shown by the parts of certain composite systems 
offer one of the most counterintuitive and subtle aspects of Quantum Mechanics (QM). 
Even if these parts are far away from each other, quantum entanglement persists,
giving rise to paradoxical situations as in the EPR configuration \cite{epr}. 
This opened the important possibility to test QM {\em vs} Local Realism (LR)
by means of Bell's inequalities \cite{bell,variousbell,clh}.
These tests should be performed in different and complementary branches 
of physics not only to avoid the loopholes
encountered in photon experiments, but also because they are 
interesting {\it per se}. 

EPR entangled pairs consisting of neutral kaons, {\it i.e.} of two massive
hadrons, are of interest at least for two reasons. One is precisely their mass, which
makes this situation quite different from the more frequently considered case of
massless photons. Another reason is the strong nature of hadronic interactions, which should
enhance detection efficiencies --rather poor for photons-- and contribute to close the
efficiency loophole. Many recent works on kaon pairs have thus appeared
\cite{ghi,eberhard,domenico,uchiyama,bf,selleri,bn,abn,BH,gigo,dg,handbook,CPLEAR}. 
One usually considers the maximally entangled state:
\begin{equation} 
\label{state}
\Phi (t) = {N(t) \over \sqrt{2}} \left[{K^0} \bar{K^0} - \bar{K^0} {K^0}\right] ,
\end{equation}
coming from $\phi$--meson decays, as at the DA$\Phi$NE $\phi$--factory 
\cite{handbook}, or proton--antiproton annihilations at rest. The latter have been 
investigated by the CPLEAR experiment \cite{CPLEAR}, confirming the 
non--separability of state (\ref{state}). In Eq.~(\ref{state}), the (proper) time dependent 
factor $N(t) \equiv e^{-i(\lambda_S +\lambda_L)t}$, with
$\lambda_{S(L)}\equiv m_{S(L)}-i\Gamma_{S(L)}/2$, accounts for kaon decays on
both sides. Apart from this, the structure of the state (\ref{state}) is identical to that
of the spin--singlet state decaying into two spin--1/2 particles, as discussed in the 
EPR--Bohm analysis, and to that of the photon entangled pair frequently used in experimental
arrangements. In all these cases one has to deal with a bipartite  system formed by two
dimension--two subsystems which fly apart along a left and a right beam. There are thus
clear and useful analogies among these various cases, as has been enphasized in
Refs.~\cite{ghi,bn,gigo}.

There is, however, an important difference which reduces considerably the possibilities
of Bell--tests with kaons. In the two spin--$1/2$ (photon) case, one can measure spin projections
(linear polarizations) on each one of the two distant subsystems along {\it any} space
direction chosen at will. 
Individual measurement outcomes are then given by a dichotomic variable
and each subsystem is projected into one of the two orthogonal states of the basis
associated to that chosen direction. The same happens when measuring the strangeness $S$
of a neutral kaon beam by forcing it to interact with dense nucleonic matter. The distinct
strong interactions of the strangeness $S=\pm 1$ states, $K^0$ and 
$\bar{K^0}$, on nucleons project the initial state into one of these two orthogonal 
members of the strangeness basis. Such a {\it strangeness measurement} is then in
complete analogy with the spin case. Alternatively, one can identify the $K_S$
{\it vs} $K_L$ components of a beam via kaon weak decays and 
strong interactions, as discussed in the following.
The $K_S$ and $K_L$ states are not strictly orthogonal, 
$\langle K_S|K_L \rangle = 2 {\cal R}e\, \epsilon / (1+|\epsilon |^2 ) \simeq 
3.2 \times 10^{-3}$, thus their identification
cannot be exact even in principle. However, the $CP$--violation parameter is so
small, $|\epsilon |\simeq 2.3 \times 10^{-3}$, and the decay probabilities of the two
components so different ($\Gamma_S \simeq 579 \Gamma_L$) that the $K_S$ {\it vs} $K_L$ 
identification can effectively work in many cases. However, apart from this (only
approximate) $K_S$ {\it vs} $K_L$ identification and the previous (in principle exact)
strangeness measurement, no other quantum--mechanical measurement
with dichotomic outcomes is possible for neutral kaons. Only these two kaon quasi--spin
directions can be used to establish Bell's inequalities. This is in sharp contrast to the spin--singlet
case and, as stated, reduces the possibilities of kaon experiments \cite{elzorro}. 

The complete analogy between strangeness measurements and spin--component 
measurements has been exploited by many authors \cite{ghi,selleri,bn,abn,gigo,dg}. 
In the analysis by Ghirardi {\it et al.}~\cite{ghi} one considers the state
(\ref{state}) produced by a $\phi$ decay and performs joint strangeness measurements at two
different times on the left beam and at other two different times on the right beam.
Because of strangeness oscillations in free space along both kaon paths, choosing
among four different times corresponds to four different choices of measurement 
directions in the spin case. In this sense, there is a total analogy and the Bell's inequalities
discussed in Ref.~\cite{ghi} are a strict consequence of LR. Unfortunately,
these inequalities are never violated by QM because strangeness
oscillations proceed too slowly and cannot compete with the more
rapid kaon weak decays. As discussed in Refs.~\cite{gigo,dg}, the inequalities exploiting
strangeness measurements at four different times can be violated by QM only if 
a normalization of the observables to undecayed kaons is employed. The 
authors of Refs.~\cite{bn,abn}, while insisting on the
convenience of performing only unambiguous strangeness measurements, have substituted
the use of different times by the possibility of choosing among different kaon regenerators 
to be inserted along the kaon path(s). 
The well known regeneration effects can be interpreted as producing rotations in the
kaon quasi--spin space analogous to the strangeness oscillations in Ref.~\cite{ghi}, without
requiring additional time intervals. One can thus derive solid Bell's inequalities,
violated by QM, for simultaneous left--right strangeness measurements.
The drawback of these analyses is that, up to now, they only refer
to thin regenerators and the predicted violations of
Bell's inequalities (below a few percent) are hardly observable. 

The alternative option, based on $K_S$ {\it vs} $K_L$ identification, has been proposed by
Eberhard \cite{eberhard}. In this case, it is convenient to 
rewrite the state (\ref{state}) as:
\begin{equation} 
\label{stateLS}
\Phi (t) \simeq  {N(t) \over \sqrt{2}} \left[ K_S K_L - K_L K_S \right] , \nonumber
\end{equation}
the small $CP$--violating effects being neglected. 
To observe if a neutral kaon in a beam is $K_S$ or $K_L$ at a given point 
({\it i.e.}~instant), a kaon detector is located far enough downstream from this point 
so that the number of undecayed $K_S$'s reaching the detector is negligible. Since 
$\Gamma_L<<\Gamma_S$, almost all $K_L$'s can reach the detector, where they
manifest by strong nuclear interactions. In a complementary way, $K_S$'s are
identified by their decays not far from that point of
interest. Misidentifications and ambiguous events will certainly appear, but at an
acceptably low level \cite{eberhard}. In Ref.~\cite{eberhard},
$K_S$ {\it vs} $K_L$ measurements are
then performed for each one of four experimental setups. In a first setup,
the state (\ref{stateLS}) is allowed to propagate in free space; its normalization is lost
because of weak decays, but its perfect antisymmetry is maintained. In the other three 
setups, thick regenerators are asymmetrically located along one beam, or along the
other, or along both. An interesting inequality relating the number of $K_L$'s detected in
each experimental setup is then derived from LR. It turns out to be
significantly violated by QM predictions even if the above mentioned detection
uncertainties are taken into account. Unfortunately, these successful predictions 
have some limitations, as already discussed by the author \cite{eberhard}. In
particular, they are valid for asymmetric $\phi$--factories (where the two neutral kaon
beams form a small angle), whose construction is not foreseen. 

The purpose of the present letter is to derive new forms of Bell's inequalities for
neutral kaons not affected by the drawbacks we have mentioned. The key point
is to combine the two kinds of dichotomic measurements, $K^0$ {\it vs} $\bar{K^0}$  and
$K_S$ {\it vs} $K_L$, in various alternative experimental setups. Like 
in Refs.~\cite{ghi,eberhard,bn,abn,gigo,dg}, where the
various measurements are either of one type or the other, one can thus  derive
inequalities which follows strictly from LR. The inequalities we obtain
turn out to be considerably violated by QM predictions, therefore a
test of LR {\it vs} QM could in principle be performed at symmetric
machines in operation.

We start with neutral kaon pairs produced in $\phi$ decays or proton--antiproton
annihilations at rest, as given by Eqs.~(\ref{state}) and (\ref{stateLS}). 
A thin regenerator is fixed, say, on the right beam very close \cite{veryclose} to the 
$\phi$ decay point. 
If the proper time $\Delta t$ required by the neutral kaon to cross the regenerator is short
enough, $\Delta t << \tau_S$, and weak decays can thus be ignored, 
$N(\Delta t) \simeq 1$, the state (\ref{stateLS}) becomes: 
\begin{equation} 
\label{state0'}
\Phi (\Delta t) \simeq {1 \over \sqrt{2}} \left[K_S K_L -K_L K_S
+r K_S K_S - rK_L K_L \right].
\end{equation}
The complex parameter $r$ characterizes the regeneration effects and is defined by
\cite{kabir}:
\begin{equation} 
r \equiv i{\pi \nu \over m_K}(f - \bar{f}) \Delta t =
i{\pi \nu  \over p_K}(f - \bar{f}) d,  \nonumber
\end{equation}
where $m_K$ is the average neutral kaon mass, $p_K$ the kaon momentum,
$f$ ($\bar{f}$) the ${K^0}$--nucleon ($\bar{K^0}$--nucleon) forward
scattering amplitude, $\nu$ the density of scattering centers of the homogeneous
regenerator whose total length is $d$.

The states (\ref{stateLS}) and (\ref{state0'}) only
differ in the terms linear in the small parameter $r$ [$|r|= {\cal O}(10^{-3})$
when $d=1$ mm \cite{domenico,kabir}]. 
To enhance their difference we now allow the state (\ref{state0'}) to propagate in free
space up to a proper time $T$, with $\tau_S << T << \tau_L$. One thus obtains the state: 
\begin{eqnarray} 
\label{stateT}
\Phi (T) &\simeq & {N(T) \over \sqrt{2}} \left[ K_S K_L - K_L K_S \right. \\
      &-& r e^{-i\Delta m T} e^{{1\over 2}(\Gamma_S - \Gamma_L)T}K_L K_L \nonumber \\
&+&\left.  r e^{i\Delta m T} e^{{1\over 2}(\Gamma_L - \Gamma_S)T}K_S K_S \right] , \nonumber
\end{eqnarray}
where $\Delta m \equiv m_L -m_S$. The $K_L K_L$ component
in (\ref{stateT}) has survived against weak decays much better than the
accompanying terms $K_S K_L$ and $K_L K_S$ and has thus been enhanced. On the contrary, 
the $K_S K_S$ component has been further suppressed and can be neglected.

The normalization of the state (\ref{stateT}) to the surviving pairs leads then to:
\begin{equation} 
\label{stateN}
\Phi = {1 \over \sqrt{2 + |R|^2}} \left[ K_S K_L  - K_L K_S 
+ R K_L K_L \right], 
\end{equation}
where:
\begin{equation} 
R \equiv  -r e^{\left[-i\Delta m +{1 \over 2}(\Gamma_S - \Gamma_L)\right]T} . \nonumber
\end{equation}
The quantity $|R| \simeq  |r| e^{ {1 \over 2}\Gamma_S T}$ is not
necessarily small with an exponential factor compensating the smallness of $|r|$. 
The state $\Phi$ is the entangled state we are going to consider for a
Bell-test. It describes all kaon
pairs with both left and right partners surviving up to a common proper time $T$. At this
point alternative measurements will be performed on each one of these kaon pairs $\Phi$.

Among the various versions of Bell's inequalities we choose to work with
that by Clauser and Horne (CH) \cite{clh}. 
For each kaon on each beam at time $T$ we have to perform either a
strangeness measurement ($K^0$ {\it vs} $\bar{K^0}$) or a lifetime measurement ($K_S$ {\it vs} $K_L$). 
Therefore, for instance, $P(K_S,\bar{K^0})$ will stay for the joint probability to observe
a $K_S$ on the left and a $\bar{K}^0$ on the right
when the appropriate experimental setup is used. 
Single--kaon probabilities can be expressed in
terms of joint probabilities and measured with the same setups. 
Therefore, $P(K_S,*) \equiv P(K_S,K^0) + P(K_S,\bar{K}^0)$ will correspond to the probability
of observing a $K_S$ on the left. In these one--side
probabilities the joint probabilities for the two possible outcomes on the other side are
added to guarantee that both kaons have survived up to time $T$.

For lifetime measurements allowing for $K_S$ {\it vs} $K_L$ identification 
at time $T$ we will apply
the following strategy. One has to detect decay events taking place in
free space between times $T$ and $T+ \Delta T$. The otherwise surviving kaons
will be detected by a dense absorber placed at the end of this decay region.  
If $ \Delta T$ is large
enough, the undecayed kaons can be identified as $K_L$'s with reasonable certainty,
while those who decayed in free space are most probably $K_S$'s. If, for illustration
purposes,  we take $ \Delta T \simeq 5.5 \tau_S$, 99.1~\% of $K_L$'s will survive, while
99.6~\% of $K_S$'s will decay. Misidentifications are thus of the order of a few
per thousand, {\it i.e.}~of the same order as the 
$CP$--violating effects we are systematically neglecting.  
Notice that being $K_S$ or $K_L$ is a stable property in free space; provided the decay is observed
between $T$ and $T+ \Delta T$, it implies that the neutral kaon was 
$K_S$ at time $T$ already. Care has to be taken, however, to choose
$T$ large enough to guarantee the space--like separation between left and right
measurements. Locality excludes then any influence from the experimental setup
encountered by one member of the kaon pair at time $T$ on the behaviour of
its other--side partner between $T$ and $T+ \Delta T$. For kaon pairs from $\phi$ decays 
moving at $\beta \simeq 0.22$ this implies $T > 1.77 \Delta T$, with a considerable
reduction of the total kaon sample. Indeed, for our previous illustrative
case one can choose $T = 2 \Delta T \simeq 11 \tau_S$, and only 1 in 60000 initial
events can be used, having both kaons surviving at $T$. 
 
Strangeness measurements at time $T$ are performed by exploiting 
the distinct $K^0$ and $\bar{K^0}$ strong interactions on nucleons. To this end, a thin but
extremely dense sheet of matter should be placed at the corresponding distance. Most of the
authors working on the subject \cite{ghi,selleri,bn,abn,gigo,dg} consider that such a
measurement is free from ambiguities and perfectly analogous to those employed in the spin
case. In principle, this is true for infinitely dense sheets behaving like
perfect absorbers and thus forcing the strong interaction to occur in a short time 
interval around $T$. In practice, the efficiency of the absorbers at disposal
is limited and complicates the issue \cite{CPLEAR}. It would be highly desirable
to identify very efficient absorbers exploiting, for instance, the formation of
baryonic resonances in low--energy $\bar{K^0}$--$N$ reactions. If this is not enough one should
proceed to introduce conventional  efficiency corrections \cite{efficiency}. In any case, the
$\bar{K^0}$--$N$ cross--sections are known to be higher than those for ${K^0}$--$N$, thus
implying a larger $\bar{K^0}$ detection efficiency. Consequently, we will express our
results in terms of $\bar{K^0}$ (rather than ${K^0}$) detection probabilities.

Notice that the two measurements described in the last two paragraphs are mutually
exclusive. One can choose to measure either $K_S$ {\it vs} $K_L$ or $K^0$ {\it vs} $\bar{K^0}$. 
In the first case no absorber is placed at time $T$ and nothing is learnt on the strangeness
of the kaon. Alternatively, if one chose to insert the absorber at time $T$ to identify    
strangeness, this is achieved only for absorbed events and nothing can be ever learnt
on the $K_S$ {\it vs} $K_L$ nature of the kaon. For
the undetected kaons appearing downstream the absorber, one fails to obtain any information
on their state at time $T$. Later they can be observed to decay soon ($K_S$'s) or not
($K_L$'s), but this information on their $K_S$ {\it vs} $K_L$ nature refers to the kaon leaving the
absorber, not to that at time $T$, which is the one of interest. 

According to the previous discussion, the requirements for deriving a Bell's inequality from LR
are fulfilled: 
\begin{itemize}
\item A non--factorizable or entangled state (\ref{stateN}) is used;
\item Alternative (mutually exclusive) measurements of either $K^0$ {\it vs} $\bar{K^0}$ or 
      $K_S$ {\it vs} $K_L$, which correspond to non--commuting observables in
      quasi--spin space, can be chosen at will; 
\item Dichotomic outcomes for each single measurement are obtained;
\item Measurement events are space--like separated.
\end{itemize}
The situation mimics perfectly that of the spin case and the very same arguments invoked by
Clauser and Horne \cite{clh} can be directly used for our present purposes. 

Avoiding to work with $K^0$ detection, because of its lower efficiency, one
can write several CH's inequalities. The most convenient ones turn out to be:
\begin{eqnarray} 
\label{CH}
-1 \leq &-&P(\bar{K^0} ,\bar{K^0} ) + P(K_S, \bar{K^0} ) + P(\bar{K^0} ,K_L)  \\
&+& P(K_S,K_L) - P(K_S ,*) - P(*,K_L) \leq 0 , \nonumber \\ 
-1 \leq &-&P(\bar{K^0} ,\bar{K^0} ) + P(\bar{K^0},K_S ) + P(K_L,\bar{K^0}) \nonumber \\
&+& P(K_L,K_S) - P(*,K_S) - P(K_L,*) \leq 0 ,  \nonumber
\end{eqnarray}
where each one follows from the other by just inverting left and right measurements on our
left--right asymmetric state (\ref{stateN}). 

As discussed in Ref.~\cite{clh}, the right--hand side (homogeneous) inequalities in 
Eq.~(\ref{CH}) have the advantage of being independent of the normalization of the total sample
of pairs involved and are thus easier to test. 
Only two things remain to be seen: does QM violate these inequalities? 
Is the predicted violation large enough to compensate the detection accuracy
level of our procedure?

The probabilities are computed in QM by just writing the state 
(\ref{stateN}) in the appropriate basis. By substituting these results in the 
homogeneous inequalities of Eq.~(\ref{CH}) one easily finds, respectively:
\begin{equation} 
\label{QM}
{2 - {\cal{R}}e\, R +{1\over 4}|R|^2 \over 2 +|R|^2} \leq 1, \; \; 
{2 + {\cal{R}}e\, R +{1\over 4}|R|^2 \over 2 +|R|^2} \leq 1, 
\end{equation}
whose only difference is the sign affecting the linear term in ${\cal{R}}e\, R$. 
According to the sign of ${\cal{R}}e\, R$, one of these two inequalities is violated if  
$|{\cal{R}}e\, R| \geq 3|R|^2/4$. The greatest violation occurs for a purely real value of $R$ 
with $|R| \simeq 0.56$, for which one of the two ratios in Eq.~(\ref{QM}) reaches the value 
1.14. This 14~\% violating effect predicted by QM is much larger than the
unavoidable inaccuracies inherent in our procedure, hence it allows, at least in
principle, for a meaningful experimental test. 

Values for the parameter $R$ satisfying $|{\cal{I}}m\, R| << |{\cal{R}}e\, R| \simeq 0.56$, as
required, are not difficult to obtain. Indeed, for kaon pairs from $\phi$ decays and according to
the values of the regeneration parameters \cite{domenico}, one can use a thin 
beryllium regenerator 1.6 mm thick to prepare the state (\ref{state0'}), which then converts into the
state (\ref{stateN}) with the desired value of $R$ by propagating in free space up to 
$T \simeq 11 \tau_S$, as previously considered. 

In a forthcoming paper \cite{elzorro} we shall discuss, systematically, the other
Bell's inequalities violated by QM that can be derived by exploiting the same
experimental setups considered here.

This work has been partly supported by the EURODAPHNE EEC--TMR 
program CT98--0169. Discussions with M. Nowakowski are also acknowledged.

\end{multicols}

\begin{references}
\bibitem{veryclose} 
Actually, it is not strictly necessary (and practically very difficult)
to place the regenerator very close 
to the $\phi$ decay point. For the following discussion
the regenerator can be placed at any instant time $t^{\prime}$
after the $\phi$ decay: the QM state is then given again by Eqs.~(\ref{state}) and
(\ref{stateLS}), once a normalization to undecayed kaon pairs at time $t^{\prime}$ is considered.
This only affects the experimental sample of counts one can then use for the Bell--test.
\end{references}
\end{document}